\documentclass[aps,prl,reprint,groupedaddress,showpacs,showkeys]{revtex4-1}

\usepackage{latexsym,amssymb,amsfonts,mathrsfs,color,graphicx,amsmath}
\usepackage[american]{babel}

\begin{document}

\newcommand{\vhh}[1]{\hat{\mathbf{#1}}}
\newcommand{\eh}{\hat{\mathbf{e}}}
\newcommand{\hsy}[1]{\hat{\boldsymbol{#1}}}
\newcommand{\RCh}{R_{\text{Ch}}}
\newcommand{\RSq}{R_{\text{Sq}}}
\newcommand{\vdd}[1]{\frac{d}{dt}\mathbf{#1}}
\newcommand{\vdh}[1]{\frac{d}{dt}\hat{\mathbf{#1}}}

\newcommand{\hphi}{\hat{\boldsymbol{\varphi}}}
\newcommand{\hrho}{\hat{\boldsymbol{\rho}}}
\newcommand{\hz}{\hat{\mathbf{z}}}
\newcommand{\ephi}{e_\varphi}
\newcommand{\erho}{e_\rho}
\newcommand{\ez}{e_z}
\newcommand{\htheta}{\hat{\boldsymbol{\theta}}}
\newcommand{\vf}{\bar{v}_f}
\newcommand{\HDD}{H_\text{2D}}
\newcommand{\HDDD}{H_\text{3D}}

\newcommand{\refEq}[1]{Eq.~(\ref{Eq:#1})}
\newcommand{\refEqu}[1]{Eqs.~(\ref{Eq:#1})}
\newcommand{\refFig}[1]{Fig.~\ref{Fig:#1}}
\newcommand{\refFigu}[1]{Figs.~\ref{Fig:#1}}
\newcommand{\refSec}[1]{Sec.~\ref{Sec:#1}}
\newcommand{\vecin}[2]{#1 \!\cdot\! #2}

\renewcommand{\phi}{\varphi}

\title{Nonlinear dynamics of a microswimmer in Poiseuille flow}
\author{Andreas Z\"{o}ttl and Holger Stark}
\date{\today}



\affiliation{Institut fur Theoretische Physik, Technische Universit\"{a}t Berlin, Hardenbergstrasse 36, 10623 Berlin, Germany}



\begin{abstract}
We study the three-dimensional dynamics of a
spherical microswimmer in cylindrical Poiseuille flow
which can be mapped onto a Hamiltonian system.
Swinging and tumbling trajectories are identified.
In 2D they are equivalent to oscillating and circling solutions of a mathematical pendulum.
Hydrodynamic interactions between the swimmer and confining channel walls lead to dissipative dynamics and result
in stable trajectories, different for pullers and pushers.
We demonstrate this behavior in the dipole approximation of the swimmer and with simulations
using the method of multi-particle collision dynamics.

\end{abstract}
\maketitle
Microswimmers often have to respond to fluid flow and confining boundaries, like
sperm cells in the Fallopian tubes \cite{Riffell07}  or pathogens in blood vessels \cite{Uppaluri11}.
Artificial microswimmers
constructed with the vision to act as  drug-deliverers in the human body \cite{Nelson10}
would have to swim in narrow channels like arteries.
Two properties influence the swimming in microchannels.
On the one hand,
vortices in flow reorient the swimming direction of microorganisms.
In simple shear flow, for example, microswimmers tumble due to a constant flow vorticity \cite{tenHagen11}.
Vortices in Poiseuille flow in combination with bottom-heaviness due to gravitation
lead to stable orientations
of swimming algae cells \cite{Pedley87}.
On the other hand, microorganisms swimming near surfaces are trapped by hydrodynamic interactions  \cite{Berke08}
and ultimately escape with the help of rotational diffusion \cite{Drescher11}.
Finally, bacteria in Poiseuille flow
show a net-upstream flux at the walls due to the interplay of confinement and flow vorticity \cite{Hill07,Nash10}.
All these examples show there is  genuine interest in understanding generic features of microorganisms and artificial 
swimmers in Poiseuille flow.

In this letter we demonstrate that the dynamics of a simple spherical microswimmer in a cylindrical Poiseuille flow can be mapped
onto a conservative dynamical system with the Hamiltonian as a constant of motion.
In analogy to the oscillating and circling solutions of a mathematical pendulum, we discuss in detail the swinging and tumbling
motion of the microswimmer in 2D and generalize them to three dimensions.
Hydrodynamic interactions with the channel wall treated in the dipole approximation
introduce \textit{dissipation} and the microswimmer assumes
specific stable swimming trajectories depending on its type as puller or pusher.


 \begin{figure}[h]
 \includegraphics[width=0.9\columnwidth]{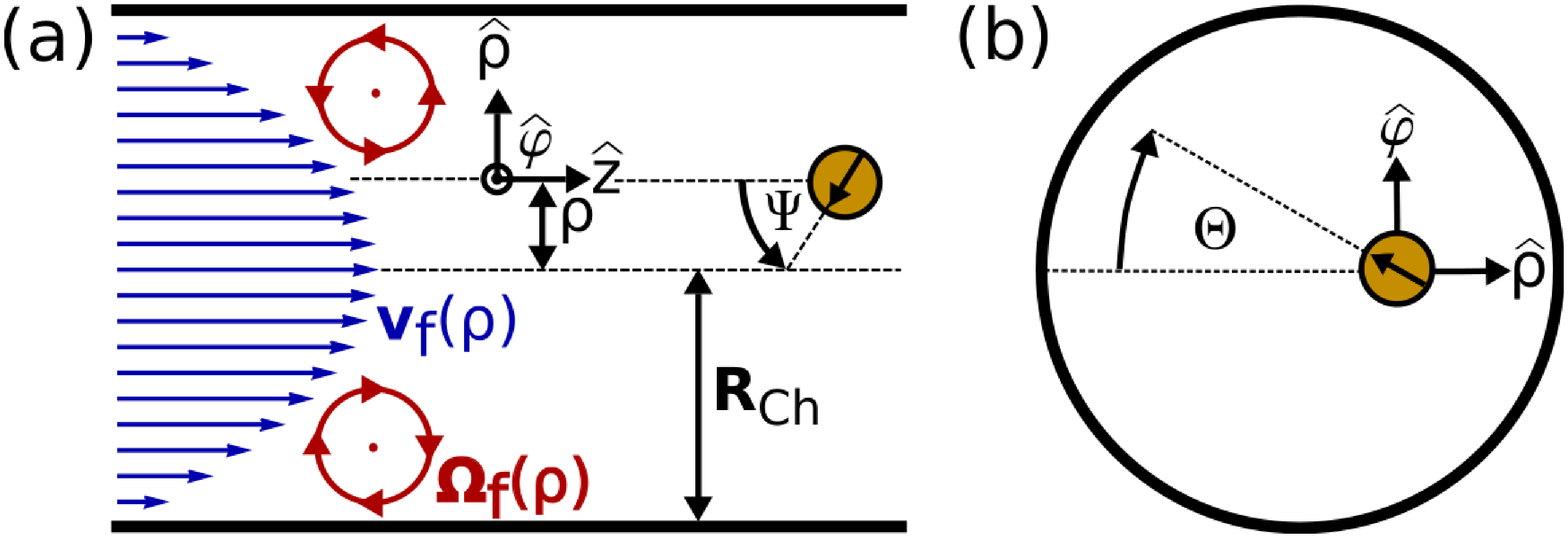}
\caption{ Swimmer in Poiseuille flow.
(a) Flow profile $\mathbf{v}_f(\rho)$, cylindrical coordinate system 
with coordinate basis ($\hrho,\hphi,\hz$)
and orientation angle $\Psi$ for the projected orientation into the  $\rho$-$z$-plane.
When $\ephi\!=\!\sin\Theta\!=\!0$, the motion is two-dimensional.
Note that the sign of vorticity $\boldsymbol{\Omega}_f$ changes when crossing the centerline.
(b) Cross section of the microchannel. The orientation in $\phi$-direction defines the angle $\Theta$.
 }
\label{Fig:Geometry}
 \end{figure}

We first introduce the geometry.
We consider a point-like microswimmer that moves with a constant intrinsic swimming speed $v_0$ in a cylindrical microchannel where a
 Poiseuille flow is imposed. Using a cylindrical coordinate system ($\rho,\phi,z$) with the coordinate basis ($\hrho,\hphi,\hz$),
the flow is given by  $\mathbf{v}_f\! =\! v_f ( 1\!-\!\rho^2 / \RCh^2 ) \vhh{z}$, 
where $v_f$ is the maximum flow speed in the center of the channel [\refFig{Geometry}(a)].
In the absence of noise the equations of motion for the swimmer position  $\mathbf{r}$ and orientation  $\eh$ are given by
\begin{equation}
\vdd{r} = v_0\eh + \mathbf{v}_f, \quad
\vdh{e} =  \frac 1 2 \boldsymbol{\Omega}_f   \times \eh
\label{Eq:EOM1}
\end{equation}
where $\boldsymbol{\Omega}_f\!=\!\nabla\! \times\! \mathbf{v}_f\! =\! v_f\rho/ \RCh^2\hphi$ is the flow vorticity.
The swimmer orientation $\eh\!=\!\erho\hrho\!+\!\ephi\hphi\!+\!\ez\hz$ has the components  
\begin{equation}
 e_{\rho} = -\cos\Theta\sin\Psi, e_{\phi}= \sin\Theta, e_z=-\cos\Theta\cos\Psi,
\label{Eq:e}
\end{equation}
where $\Psi\!\in\! \{-\pi,\pi\}$  is the angle in the $\rho$-$z$-plane  [\refFig{Geometry}(a)] and
$\Theta\!\in\! \{-\pi/2,\pi/2\}$  measures the orientation in the azimuthal $\phi$-direction [\refFig{Geometry}(b)].
We note that $|\Psi|\!<\! \pi/2$ means upstream and $|\Psi|\!>\! \pi/2$  downstream orientation,  respectively.
In the following we use rescaled units, $\rho/ \RCh\!\rightarrow\!\rho\!\in\! \{0,1\}$,
$z/ \RCh\!\rightarrow\!z$ and $t/t_0\!\rightarrow\!t $ with $t_0 \!=\!  \RCh / v_0$.
We also introduce the dimensionless flow speed $\vf\!=\!v_f/v_0$, which is the only essential parameter in our problem.

\begin{figure}[h]
\includegraphics[width=\columnwidth]{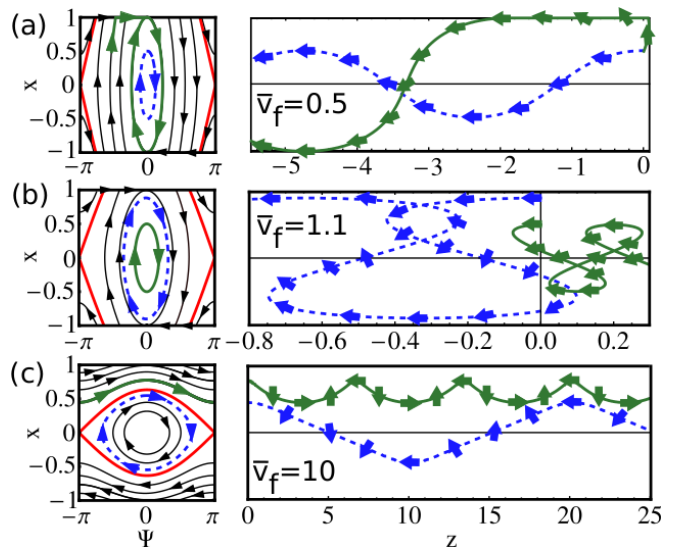}
\caption{
Phase spaces $x$-$\Psi$ (left) and typical trajectories $z(x)$ (right) for several flow strengths $\vf$.
All trajectories start at $z\!=\!0$.
(a) upstream motion,
(b) intermediate motion and
(c) downstream motion.
Note the various scales for the z-axis.
The arrows indicate the orientation vector $\eh$ of the swimmer.
 }
\label{Fig:2D}
 \end{figure}

First, we discuss 2D solutions of Eqs.~(\ref{Eq:EOM1}) since
they already capture many aspects of the swimmer dynamics.
When $\Theta\!=\!0$, the trajectories of the swimmer
are restricted to 
two dimensions, for example, to the $x$-$z$-plane,   $x\!\in\!\{-1,1\}$.  
Due to the translational symmetry in $z$-direction, only the equations for  $x$ and $\Psi$ are coupled, 
and \refEqu{EOM1} give
$ \dot{x}\! =\! -\sin\Psi $, $ \dot{\Psi}\! =\! \bar{v}_fx $.
Eliminating $x$ results in
\begin{equation}
 \ddot{\Psi} = \bar{v}_f\sin\Psi,
 \label{Eq:Pend1}
\end{equation}
which is the equation of motion of the mathematical pendulum. 
Since in this analogy $x$ plays the role of velocity, we can immediately write down the
2D Hamiltonian
\begin{equation}
   \HDD=\frac 1 2 \vf x^2 +1-\cos\Psi
 \label{Eq:H2D}
\end{equation}
as a conserved quantity.
\refFig{2D} shows the $x$-$\Psi$ phase space and
typical trajectories $z(x)$ for several flow strengths $\vf$.
In analogy to the pendulum two swimming states exist.
The flow vorticity rotates the upstream oriented microswimmer always towards the center.
Hence, the swimmer
performs a swinging motion around the centerline of the channel for $\HDD\!<\!2$
which corresponds to the oscillating solution of the pendulum [e.~g.~blue trajectory of \refFig{2D}(a)].  
For small amplitudes ($\Psi\! \ll\! 1$) the swinging frequency is $\omega_0\!=\!\sqrt{\vf}$. 
When the upstream oriented swimmer moves exactly in the center of the channel (stable fixed point),  
the Hamiltonian is zero.
Downstream swimming along the centerline ($\Psi\!=\!\pi$) is
an unstable fixed point. 
After a slight disturbance of $x\!=\!0$, vorticity rotates the swimmer away from the centerline.
The swimmer performs a tumbling motion ($\HDD\!>\!2$) which corresponds to the circling solution
of the pendulum [green trajectory of \refFig{2D}(c)].
At $\HDD\!=\!2$,
the separatrix $x^2\!=\!2(\cos\Psi\!+\!1)/ \vf$ divides the swinging and tumbling region in phase space [red curves in the phase portraits
of \refFig{2D}].
Since the Poiseuille flow is bounded by the channel walls,  
tumbling motion only occurs for $\vf\!>\!4$ [\refFig{2D}(c)].
Meaning sufficiently strong vorticity
to prevent the swimmer from crossing the centerline.

If we only consider steric interactions of the swimmer with the channel wall,
the  swimmer crashes into the wall at $|x|\!=\!1$ for $\HDD\!>\!\vf/2$, reorients due to the flow vorticity towards the upstream orientation,
and leaves the wall at $\Psi\!=\!0$ with $\HDD^{\text{max}}\!=\!\vf/2$.
The swimmer then performs a swinging motion between the walls with maximum amplitude $|x|\!=\!1$
for  $\vf\!<\!4$ [green trajectory in \refFig{2D}(a)].  
So for  $\vf\!<\!4$ the swimmer always enters a swinging motion oriented upstream, at the latest after contact with the wall,
whereas it tumbles close to the wall for $\vf\!>\!4$.

To determine the full 2D trajectory in the microchannel,
we solve the dynamic equation for $z(t)$, 
\begin{equation}
 \dot{z} = \vf [1-x(t)^2] - \cos\Psi(t).
 \label{Eq:Pendz}
\end{equation}
A careful analysis reveals the following.
The swimmer always moves upstream ($\dot{z}\!<\!0$), when
 $\vf\!<\!1\!-\!\HDD$, as shown in \refFig{2D}(a).
When the flow is strong ($\vf\!>\!1\!+\!2\HDD$), the swimmer always
drifts downstream ($\dot{z}\!>\!0$), while swinging or tumbling   [\refFig{2D}(c)].
In between,
mixed up- and downstream segments within one trajectory [\refFig{2D}(b)] exist 
but a net upstream motion only occurs for $\vf\! \lesssim\! 1\!+\!\HDD/2$ [blue line in \refFig{2D}(b)].

For a non-zero azimuthal component, $e_\phi\! =\! \sin \Theta\! \neq\! 0$, the swimmer trajectory is three-dimensional.
Using \refEqu{EOM1} and   \refEqu{e},
we obtain three coupled equations for  $\Psi$, $\Theta$ and $\rho$,
 \begin{equation}
    \begin{split}
      \dot{\rho} &= -\cos\Theta\sin\Psi\\
      \dot{\Psi} &= \bar{v}_f\rho - \sin\Theta \tan\Theta\cos\Psi / \rho \\
      \dot{\Theta} &= \sin\Theta\sin\Psi / \rho .
      \label{Eq:EOM3a}
    \end{split}
  \end{equation}
Due to translational symmetry in $z$-direction and rotational symmetry about the channel axis,
\refEqu{EOM3a} do not depend on $z$ and $\varphi$.
We are able to identify two constants of motion,
\begin{equation}
    \begin{split}
      L_z &= \rho\sin\Theta \\
      H &=   \frac 1 2 \bar{v}_f \rho^2 + 1 -  \cos\Psi\cos\Theta, 
      \label{Eq:EOM3b}
    \end{split}
  \end{equation}
where $L_z$ is proportional to the angular momentum of the swimmer in $z$-direction.
Due to this constant the sign of $\Theta$ along a swimmer trajectory does not change.
Eliminating $\Theta$ from \refEqu{EOM3a},  reduces the
equations of motions to $ \frac{\partial H}{\partial \Psi}\! =\!  -\dot{\rho}$, 
    $\frac{\partial H}{\partial \rho}\! =\!  \dot{\Psi}$.
So, $H$ again plays the role of a Hamiltonian for the conjugate variables $\Psi$ and $\rho$.

\begin{figure}[h]
 \includegraphics[width=0.9\columnwidth]{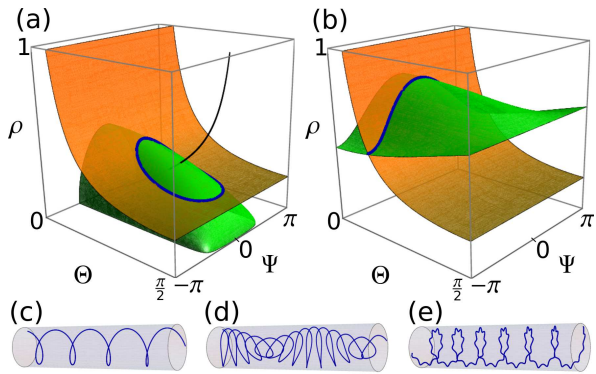}
 \caption{\label{Fig:vf2} $\rho$-$\Psi$-$\Theta$ phase space. 
 The intersection between $L_z\!=\!\text{const.}$ (orange) and $H\!=\!\text{const.}$ (green)
 gives the phase space trajectory.
 (a) helical-like swinging motion (blue intersection curve)
 for $L_z\!=\!0.2$, $H\!=\!1$.
 Black curve: fixed-point line  
 corresponds to helical trajectories.
 (b)  helical-like tumbling motion (blue intersection curve)
 for $L_z\!=\!0.2$, $H\!=\!3$.
 (c-e)  sketch of trajectories in the channel
 for helical motion (c), helical-like motion (d) 
 and tumbling motion (e).
} 
 \end{figure}

The intersection of the two constants of motion gives the orbit of the swimmer in $\rho$-$\Psi$-$\Theta$
phase space [\refFig{vf2}(a,b)].
The stable fixed points of \refEqu{EOM3a} lie on the fixed-point line
  ($ \rho^\ast\! =\!\sqrt{\sin\Theta^\ast\tan\Theta^\ast / \vf}$, $  \Psi^\ast\!=\!0$),
drawn in \refFig{vf2}(a).
Swimming at a fixed point corresponds to a helical trajectory [\refFig{vf2}(c)].
The swimmer moves upstream for $\vf\!<\!1/\cos\Theta^\ast$, either
on a left-handed helix ($\Theta\!>\!0$) or a right-handed helix ($\Theta\!<\!0$).
Closed orbits around the fixed-point line correspond to swinging motion  around a helical path [\refFig{vf2}(d)]
and open orbits [\refFig{vf2}(b)] are complicated tumbling trajectories [\refFig{vf2}(e)].

Now we consider hydrodynamic interactions of the microswimmer with the bounding channel wall. 
The flow field for neutrally buoyant swimmers in a bulk fluid is in leading order 
a force dipole, $\mathbf{v}(\mathbf{r})\! =\! \frac{p}{8\pi\eta r^2}[3(\vhh{r}\!\cdot\!\eh)^2\!-\!1]\vhh{r}$
where $\vhh{r}\!=\!\mathbf{r}/r$ and $\eta$ is the viscosity of the fluid.
For positive dipole strength, $p\!>\!0$, the propelling apparatus of the swimmer is typically at the back (\textit{pusher}),
and for $p\!<\!0$ in the front (\textit{puller}). 
Ref.~\cite{Berke08} treated the swimmer close to a plane wall.
The authors showed
that hydrodynamic interactions between the swimmer and the wall lead to
re-orientation due to the wall-induced vorticity $\boldsymbol{\Omega}_W$  
and to attraction/repulsion. 
For example, a pusher swimming parallel to the wall is attracted and remains trapped at the wall,
while a puller is repelled from the wall.
We note that close to a wall the force-dipole approximation for $\boldsymbol{\Omega}_W$
is no longer valid.
Instead, geometric details of the swimmer and thereby near fields
become important for $\boldsymbol{\Omega}_W$.
Combining flow and wall effects, the total angular velocity of a swimmer near a wall is
$\boldsymbol{\Omega}(x,\Psi)\!=\!\boldsymbol{\Omega}_W(x,\Psi)\!+\!\boldsymbol{\Omega}_f(x)$.
Recent work showed 
that hydrodynamic re-orientation is almost negligible for bacteria swimming \textit{near} walls \cite{Li09,Drescher11}
but rotates the swimmer \textit{at} the wall.
For a sufficiently weak Poiseuille flow field,
stable orientations $\Psi_W$ for swimming at the wall exist when
$\boldsymbol{\Omega}_W(\RCh,\Psi_W)\!+\!\boldsymbol{\Omega}_f(\RCh)\!=\!0$ \cite{Zoettl11}.
Thermal fluctuations will, however, reorient the swimmer so that it leaves the wall \cite{Drescher11}.

\begin{figure}[h]
\includegraphics[width=0.9\columnwidth]{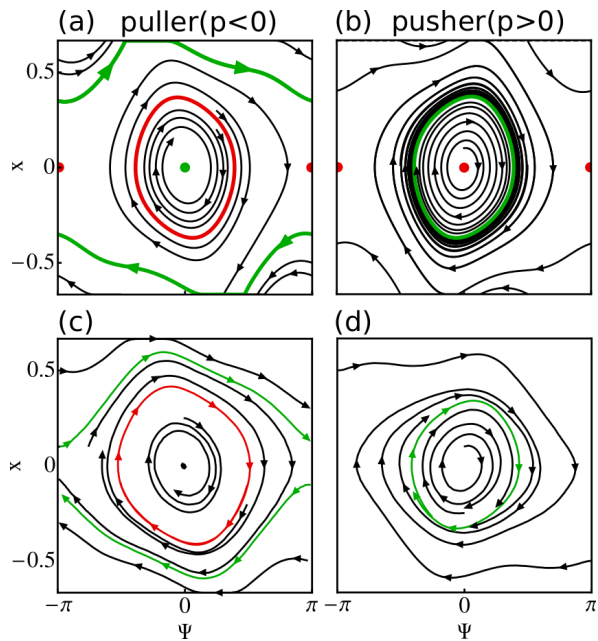}
 \caption{Phase space trajectories for
 a microswimmer in a narrow channel for $\vf\!=\!10$ for a puller (left)
 and a pusher (right).
Green and red indicate, respectively, stable and unstable trajectories.
 (a) and (b) are obtained from \refEqu{HI}
 and (c) and (d) from MPCD simulations
 where we used the parameters
 $B_1\!=\!0.045$, $\beta\!=\!B_2/B_1\!=\!\pm 5$, $R_S\!=\!6$,
 $\RCh\!=\!18$, $N_c\!=\!30$, $\Delta t\!=\!0.02$ setting $a\!=\!m\!=\!k_BT\!=\!1$
 and the collision rule MPC-AT-a \cite{Noguchi08}.
 Each trajectory was obtained by averaging over 10 individual runs.
 }
 \label{Fig:narrow}
 \end{figure}

In narrow channels a microswimmer experiences hydrodynamic interactions with the wall
all the time.
To capture the basic idea, we concentrate on 2D-trajectories and consider instead of the cylindrical channel wall,
two parallel plates located at $x\!=\!1$ and $x\!=\!-1$. 
We calculate the wall-induced translational and angular velocities using
the force-dipole approximation of Ref.~\cite{Berke08} and obtain the
equations of motion,
\begin{equation}
 \begin{split}
 \dot{x}&=-\sin\Psi - \frac{3\bar{p}(3\sin^2\Psi-1)}{64\pi} \left(\frac{1}{(1-x)^2}-\frac{1}{(1+x)^2}\right) \\
  \dot{\Psi}&=\vf x - \frac{3\bar{p}\sin\Psi\cos\Psi}{64\pi}\left(\frac{1}{(1-x)^3}+\frac{1}{(1+x)^3}\right)
 \end{split}
\label{Eq:HI}
\end{equation}
where $\bar{p}\!=\!p / ( \eta v_0)$ is the reduced dipole strength.
\refFig{narrow} shows typical phase space plots generated from \refEqu{HI} for a puller (a)
and pusher (b).
The swinging motion of an upstream oriented puller becomes damped
and an attractive fixed point in the center exists.
Repelled by both walls, the puller swims upstream along the centerline.
On the other hand, a puller tumbling near the wall is attracted, on average, by the wall and stays near to it.
This is indicated by the green stable trajectory.
All trajectories outside the unstable red limit cycle or separatrix converge to it.
However, due to thermal fluctuations the puller may cross the separatrix.
 A   pusher behaves differently, it
  is attracted by the wall when oriented upstream in the center of the channel,
  but it is pushed away from the wall when tumbling near the wall.
So, all trajectories converge towards a swinging motion about the centerline,
  characterized by a stable limit cycle in the $x$-$\Psi$-plane.

To test our findings we simulate the motion of a spherical microswimmer in Poiseuille flow using the method of
multi-particle collision dynamics (MPCD) \cite{Malevanets99}.
It solves the Navier-Stokes equations on a coarse-grained level and calculates the flow field
around the swimmer in the cylindrical microchannel taking into account both hydrodynamic interactions and thermal noise.
In every simulation step randomly distributed point-particles of mass $m$ at temperature $k_BT$
first move ballistically for a time $\Delta t$ and then they are sorted into
cubic cells of length $a$.
They interact with all other particles in the cell with a specific
collision rule such that momentum is conserved locally.
The density of the fluid is $mN_c/a^3$ where $N_c$ is the average number
of particles per cell.
Depending on the parameters and the specific collision rule, the viscosity $\eta$ of the fluid can be calculated \cite{Malevanets99,Noguchi08}.

As a model microswimmer we use a spherical squirmer of radius $R_S$   \cite{Squirmer,Ishikawa06}.
It propels itself by a static,
axisymmetric and tangential  velocity field on its surface, 
$
  \mathbf{v}_s(\hat{\mathbf{r}}_s,\eh)\!=\!
  \left( B_1\! +\! (\vecin{\eh}{\hat{\mathbf{r}}_s})B_2    \right)  \left[ (\vecin{\eh}{\hat{\mathbf{r}}_s})\mathbf{\hat{\mathbf{r}}_s} \!-\!\eh  \right]
$
where $\hat{\mathbf{r}}_s$ is the radial unit vector pointing from the center of the squirmer to the surface. 
The first mode $B_1$ determines the swimming speed $v_0\!=\!2B_1/3$ and the second mode
the strength of the force dipole $p/ \eta\! =\! -4\pi B_2R_S^2$, so $B_2\!>\!0$ models a puller and $B_2\!<\!0$ a pusher.
The squirmer has already been used to investigate hydrodynamic interactions between several swimmers \cite{Ishikawa06,Goetze10}  and
between a swimmer and a wall \cite{Llopis10}
and has been realized in experiments  quite recently \cite{Thutupalli11}.
To implement the squirmer in the MPCD fluid, we follow previous work \cite{Downton09,Goetze10}.
\refFigu{narrow}(c)-(d) show simulated trajectories for several initial conditions for a puller (c) 
and a pusher (d). Our parameters are listed in the caption of \refFig{narrow}.
Although near field effects and the large extent of the swimmer 
play a role in the simulated dynamics,
the qualitative behavior arising from  hydrodynamic interactions between the swimmer and the cylindrical channel wall
follows the analytical model.

In order to learn about typical relaxation times
in which swimmers approach their stable trajectories, 
we linearize \refEqu{HI} around the fixed point in the center.
We obtain a harmonic oscillator equation
for $\Psi$
with a \textsl{friction coefficient} $\gamma$ linear
in the dipole strength, 
\!$\gamma\!=\!- 3p/ (64\pi\eta \RCh^3)  $.
For the squirmer, the estimated relaxation time for swinging motion becomes
$\gamma^{-1}\!=\!t_0\!\cdot\!32\bar{R}^3/(9\beta)\!$, where $\beta\!=\!B_2/B_1$, $\bar{R}\!=\!\RCh/ R_S$,
and  $t_0 \!=\!  \RCh / v_0\!\approx\!1\,s$ is a characteristic time scale for narrow microchannels.
Typical values for $\beta$ range from $-1$ to $+1$ for existing microswimmers \cite{Evans11}.
So experiments should be able to observe that microswimmers approach their stable trajectories within seconds in sufficiently narrow channels.
Similar estimates apply to the \textit{E.~coli} bacterium
where $p\! \approx\! 0.8 pN \mu m$ was measured recently \cite{Drescher11}.
Although hydrodynamic interactions between a small microswimmer and a single  wall may play no significant role,
they become important in channels when the channel diameter  is only few times the size of the swimmer.

In conclusion, through a formal mapping onto a Hamiltonian dynamical system
we have shown that spherical microswimmers
perform either an upstream oriented swinging or 
a tumbling motion when moving in Poiseuille flow.
Hydrodynamic interactions of the swimmer with the wall stabilizes the upstream orientation of pullers
in the center of the channel
whereas a pusher performs stable oscillations around the centerline with a specific amplitude.

Spherical artificial swimmers
with different locomotion mechanisms
have been constructed 
 and studied recently \cite{swimm}.
Investigating them in microfluidic channels under Poiseuille flow, the generic features presented in this article should be accessible in experiments.

We thank R.~Goldstein,  I.~Pagonabarraga, T.~Pfohl, S.~Uppaluri and  R.~Vogel
for helpful discussions and the Deutsche Forschungsgemeinschaft for financial support
through the research training group GRK1558.


\begin{thebibliography}{10}

\bibitem{Riffell07}
J.~A.~Riffell  and R.~K.~Zimmer,
J.\ Exp.\ Biol.\ \textbf{210}, 3644 (2007).

\bibitem{Uppaluri11}
S.~Uppaluri et al.\ (unpublished).

\bibitem{Nelson10}
B.~J.~Nelson, I.~K.~Kaliakatsos, and J.~J.~Abbott,
Annu.\ Rev.\ Biomed.\ Eng.\ \textbf{12}, 55 (2010).

\bibitem{tenHagen11}
B.~ten Hagen, R.~Wittkowski, and H.~L\"{o}wen,
Phys.\ Rev.\ E \textbf{84}, 031105 (2011).


\bibitem{Pedley87}
J.~O.~Kessler, Nature (London) \textbf{313}, 218 (1985).

\bibitem{Berke08}
A.~P.~Berke, L.~Turner, H.~C.~Berg, and E.~Lauga,
Phys.\ Rev.\ Lett.\ \textbf{101}, 038102 (2008).

\bibitem{Hill07}
J.~Hill, O.~Kalkanci, J.~L.~McMurry, and H.~Koser,
Phys.\ Rev.\ Lett.\ \textbf{98}, 068101 (2007). 

\bibitem{Nash10}
R.~W.~Nash, R.~Adhikari, J.~Tailleur, and M.~E.~Cates,
Phys.\ Rev.\ Lett.\ \textbf{104}, 258101 (2010).

\bibitem{Li09}
G.~Li and J.~X.~Tang, Phys.\ Rev.\ Lett.\ \textbf{103}, 078101 (2009).

\bibitem{Drescher11}
K.~Drescher, J.~Dunkel, L.~H.~Cisneros, S.~Ganguly, and R.~E.\ Goldstein,
Proc.~Natl.~Acad.~Sci.~USA \textbf{108}, 10940 (2011).

\bibitem{Zoettl11}
A.~Z\"{o}ttl and H.~Stark (unpublished).

\bibitem{Malevanets99}
A.~Malevanets and R.~Kapral, J.\ Chem.\ Phys.\ \textbf{110}, 8605 (1999).

\bibitem{Squirmer}
J.~Lighthill, Commun.\ Pure Appl.\ Math. \textbf{5}, 109 (1952);
J.~R. Blake, J.\ Fluid Mech.\ \textbf{46}, 199 (1971).

\bibitem{Ishikawa06}
T.~Ishikawa, M.~P.~Simmonds, and T.~J.~Pedley,
J.\ Fluid Mech.\ \textbf{568}, 119 (2006).

\bibitem{Goetze10}
I.~O.~G\"otze and G.~Gompper,
Phys.\ Rev.\ E \textbf{82}, 041921 (2010).

\bibitem{Llopis10}
I.~Llopis and I.~Pagonabarraga,
J.\ Non-Newtonian.\ Fluid Mech.\ \textbf{165}, 946 (2010).

\bibitem{Thutupalli11}
S.~Thutupalli, R.~Seemann, and S.~Herminghaus,
New J.\ Phys.\ \textbf{13}, 073021 (2011).

\bibitem{Noguchi08}
H.~Noguchi, and G.~Gompper,
Phys.\ Rev.\ E \textbf{78}, 016706 (2008).

\bibitem{Downton09}
M.~T.~Downton and H.~Stark,
J.\ Phys.\ Condens.\ Matt.\ \textbf{21}, 204101 (2009).

\bibitem{Evans11}
A.~A.~Evans, T.~Ishikawa, T.~Yamaguchi, and E.~Lauga,
Phys.\ Fluids \textbf{23}, 111702 (2011).

\bibitem{swimm}
See, e.~g., 
J.~R.~Howse et al., Phys.\ Rev.\ Lett.\ \textbf{99}, 048102 (2007);
J. Palacci et al.,  Phys.\ Rev.\ Lett.\ \textbf{105}, 088304 (2010);
H.-R.~Jiang et al., Phys.\ Rev.\ Lett.\ \textbf{105}, 268302 (2010).


\end{thebibliography}
\end{document}